\documentclass[]{article}
\usepackage{lmodern}
\usepackage{amssymb,amsmath}
\usepackage{ifxetex,ifluatex}
\usepackage{fixltx2e} % provides \textsubscript
\ifnum 0\ifxetex 1\fi\ifluatex 1\fi=0 % if pdftex
  \usepackage[T1]{fontenc}
  \usepackage[utf8]{inputenc}
\else % if luatex or xelatex
  \ifxetex
    \usepackage{mathspec}
  \else
    \usepackage{fontspec}
  \fi
  \defaultfontfeatures{Ligatures=TeX,Scale=MatchLowercase}
\fi
\IfFileExists{upquote.sty}{\usepackage{upquote}}{}
\IfFileExists{microtype.sty}{%
\usepackage{microtype}
\UseMicrotypeSet[protrusion]{basicmath} % disable protrusion for tt fonts
}{}
\usepackage[unicode=true]{hyperref}
\hypersetup{
            pdftitle={(Lack Of) Representation of Non Western World in process of creation of Web standards},
            pdfauthor={Harsh Gupta},
            pdfkeywords={W3C, diversity, web standards},
            pdfborder={0 0 0},
            breaklinks=true}
\usepackage{longtable,booktabs}
\IfFileExists{parskip.sty}{%
\usepackage{parskip}
}{% else
\setlength{\parindent}{0pt}
\setlength{\parskip}{6pt plus 2pt minus 1pt}
}
\ifx\paragraph\undefined\else
\let\oldparagraph\paragraph
\renewcommand{\paragraph}[1]{\oldparagraph{#1}\mbox{}}
\fi
\ifx\subparagraph\undefined\else
\let\oldsubparagraph\subparagraph
\renewcommand{\subparagraph}[1]{\oldsubparagraph{#1}\mbox{}}
\fi

\title{(Lack Of) Representation of Non Western World in process of creation of
Web standards}
\author{Harsh Gupta}
\date{}

\begin{document}
\maketitle
\begin{abstract}
World Wide Consortium (W3C) as an standard setting organization for the
world wide web plays a very important role in shaping the web. We focus
on the ongoing controversy related to Encrypted Media Extensions (EME)
and found that there was a serious lack of participation from people
from non western countries. We also found serious lack of gender
diversity in the EME debate.
\end{abstract}

W3C is the organization which sets the standard for HTML 5. Recently it
got surrounded by controversy due to the Encrypted Media Extensions
(EME) draft specification (David Dorwin et al. 2016). EME aims to
prevent piracy of digital video by making it hard to download the
unencrypted video stream. But it also raises lots of issues regarding
implementation in Free and Open Source Software, Interoperability,
Privacy, Security, Accessibility and fair use. (Cory Doctorow 2016)

In this study we looked at aspects of the debate which both of the sides
ignored, the third world! We found that out of 48 people who
participated in the debate around EME on W3C's public-html mailing list,
none of them were from the continents of Asia, Africa or South America.
These regions make up almost 80 \% of the world's population and more
than 60 percent of world's internet users (Stats 2016). When a group of
people doesn't get represented a in the standard making process it is
expected that their concerns don't get represented either. The
representation of people is specially important in the EME debate
because laws around Digital Rights Management around the world are
different. For example Indian laws does not disallow manufacture and
distribution of circumvention tools whereas the law in USA does (Prakash
2016b). The cultural norms around the world are quite different and also
the conditions under which people use the internet are different. India
has the lowest average internet speed across the world (Akamai 2016). A
large of fraction of Indian population (37\% in 2010) accesses internet
through Cyber Cafés (TRAI 2016). These factors makes the ability to
download digital content much more important for an Indian internet user
than a North American or European internet user.

\section{Methodology}\label{methodology}

We used \texttt{BigBang}\footnote{https://github.com/datactive/bigbang}
python package to download the achieves of the \texttt{public-html}
mailing list at W3C.\footnote{https://lists.w3.org/Archives/Public/public-html/}
Our dump contains all the messages between 31st August 2010 to 15th May
2016. Then we filtered out all the emails with \texttt{EME},
\texttt{encrypted\ media} or \texttt{DRM} in the subject line. There
were 472 such emails. We then de-duplicated the list of senders as some
senders used multiple emails in the course of discussion. There were 48
unique senders afters de duplication. Then we looked up their social
media profiles (LinkedIn, Twitter, Github), personal website or page at
employers site to determine the region they belong to and their gender.
All the source code used for the analysis is available on our github
repository.\footnote{https://github.com/hargup/eme\_diversity\_analysis}

\section{Result}\label{result}

\begin{longtable}[]{@{}lll@{}}
\caption{Regional Diversity}\tabularnewline
\toprule
Region & Participant (\%) & Email (\%)\tabularnewline
\midrule
\endfirsthead
\toprule
Region & Participant (\%) & Email (\%)\tabularnewline
\midrule
\endhead
Africa & 0 (0) & 0 (0)\tabularnewline
Asia & 0 (0) & 0 (0)\tabularnewline
Australia and New Zealand & 5 (10.4) & 16 (3.4)\tabularnewline
Europe & 13 (27.1) & 146 (30.9)\tabularnewline
North America & 30 (62.5) & 310 (65.7)\tabularnewline
South America & 0 (0) & 0 (0)\tabularnewline
----------------------------- & ------------------ &
----------\tabularnewline
Total & 48 (100) & 472 (100)\tabularnewline
\bottomrule
\end{longtable}

As mentioned in the Introduction above there was absolutely no
participation from the whole continents of Africa, Asia, or South
America with most of the emails being sent by North Americans.

\begin{longtable}[]{@{}lll@{}}
\caption{Gender Diversity}\tabularnewline
\toprule
Gender & Participant(\%) & Email(\%)\tabularnewline
\midrule
\endfirsthead
\toprule
Gender & Participant(\%) & Email(\%)\tabularnewline
\midrule
\endhead
Male & 47 (97.9) & 466 (98.7)\tabularnewline
Female & 1 (2.1) & 6 (1.3)\tabularnewline
----------------------------- & ------------------ &
----------\tabularnewline
Total & 48 (100) & 472 (100)\tabularnewline
\bottomrule
\end{longtable}

The debate inside W3C around EME also seriously lacked in gender
diversity, which is typical of open communities.

\section{Discussion and Future Work}\label{discussion-and-future-work}

The lack of diversity in W3C is not unique. (Graham, Straumann, and
Hogan 2015) showed a significant western bias in Wikipedia, gender bias
in Wikipedia has also a well known and is being actively worked upon.
Internet Corporation for Assigned Names and Numbers has also been
criticized for under representing interests of non North American and
West European world (Prakash 2016a).

We hope that W3C and other organizations will increase the diversity in
their standard making process so that global voices actually shape the
global internet.

\section{Acknowledgement}\label{acknowledgement}

This work was done during my internship at The Center for Internet \&
Society, India. I thank Sunil Abraham for useful and timely feedback and
Pranesh Prakash, Amber Sinha and Udbhav Tiwari for informed discussions.

\section*{References}\label{references}
\addcontentsline{toc}{section}{References}

\hypertarget{refs}{}
\hypertarget{ref-akamai_q1_2016}{}
Akamai. 2016. ``Akamai State of the Internet Q1 2016.'' Accessed August
20.
\url{https://www.akamai.com/uk/en/multimedia/documents/state-of-the-internet/akamai-state-of-the-internet-report-q1-2016.pdf}.

\hypertarget{ref-doctorow_interoperability_2016}{}
Cory Doctorow. 2016. ``Interoperability and the W3C: Defending the
Future from the Present.'' \emph{Electronic Frontier Foundation}.
\url{https://www.eff.org/deeplinks/2016/03/interoperability-and-w3c-defending-future-present}.

\hypertarget{ref-eme_draft}{}
David Dorwin, Jerry Smith, Mark Watson, and Adrian Bateman. 2016.
``Encrypted Media Extensions, W3C Editor's Draft.'' Accessed May 13.
\url{https://w3c.github.io/encrypted-media/}.

\hypertarget{ref-graham_mapping_wikipedia}{}
Graham, Mark, Ralph K. Straumann, and Bernie Hogan. 2015. ``Digital
Divisions of Labor and Informational Magnetism: Mapping Participation in
Wikipedia.'' \emph{Annals of the Association of American Geographers}
105 (6): 1158--78.
doi:\href{https://doi.org/10.1080/00045608.2015.1072791}{10.1080/00045608.2015.1072791}.

\hypertarget{ref-pranesh_icann}{}
Prakash, Pranesh. 2016a. ``CIS Statement at ICANN 49's Public Forum.''
\emph{The Centre for Internet and Society}. Accessed August 20.
\url{http://cis-india.org/internet-governance/blog/icann49-public-forum-statement}.

\hypertarget{ref-pranesh_tpm}{}
---------. 2016b. ``Technological Protection Measures in the Copyright
(Amendment) Bill, 2010.'' \emph{The Centre for Internet and Society}.
Accessed August 20.
\url{http://cis-india.org/a2k/blogs/tpm-copyright-amendment}.

\hypertarget{ref-internet_live_stats}{}
Stats, Internet Live. 2016. ``Number of Internet Users (2016) - Internet
Live Stats.'' Accessed August 20.
\url{http://www.internetlivestats.com/internet-users/}.

\hypertarget{ref-trai_bb_reco}{}
TRAI. 2016. ``Recommendations on National Broadband Plan.'' Accessed
August 20.
\url{http://www.trai.gov.in/WriteReadData/Recommendation/Documents/Rcommendation81210.pdf}.

\end{document}